\documentclass[%
reprint,
superscriptaddress,
showkeys,
 amsmath,amssymb,
 aps,
floatfix,
]{revtex4-2}
\usepackage{mathrsfs}
\usepackage{amsmath}
\usepackage{bbold}
\usepackage{graphicx}
\usepackage{dcolumn}
\usepackage{bm}
\usepackage{upgreek}
\usepackage{lineno}
\usepackage{lmodern}
\usepackage{amsmath}
\usepackage{amssymb}
\usepackage{ulem}

\usepackage{cfr-lm}
\usepackage{color,soul}
\usepackage{upgreek}
\usepackage{mathastext}
\usepackage{setspace}
\usepackage{eqnarray,amsmath}
\usepackage{setspace}

\begin{document}

\preprint{APS/123-QED}


\title{\centering Wave momentum shaping for moving objects in heterogeneous and dynamic media}

\author{Bakhtiyar Orazbayev}
\affiliation{%
 Laboratory of Wave Engineering, \'Ecole Polytechnique F\'ed\'erale de Lausanne, Switzerland.
}%
\affiliation{%
Department of Physics, Nazarbayev University, Astana, Kazakhstan.
}%
\author{Matthieu Malléjac}
\affiliation{%
 Laboratory of Wave Engineering, \'Ecole Polytechnique F\'ed\'erale de Lausanne, Switzerland.
}%
\author{Nicolas Bachelard}
\affiliation{%
 Université de Bordeaux, CNRS, LOMA, UMR 5798, F-33405, Talence, France.
}%
\author{Stefan Rotter}
\affiliation{Institute of Theoretical Physics, Vienna University of Technology (TU Wien), Vienna, Austria}
\author{Romain Fleury}%
 \email{romain.fleury@epfl.ch}
 \affiliation{%
 Laboratory of Wave Engineering, \'Ecole Polytechnique F\'ed\'erale de Lausanne, Switzerland.
}%


\date{\today}

\begin{abstract}
Light and sound waves have the fascinating property that they can move objects through the transfer of linear or angular momentum. This ability has led to the development of optical and acoustic tweezers, with applications ranging from biomedical engineering to quantum optics. Although impressive manipulation results have been achieved, the stringent requirement for a highly controlled, low-reverberant, and static environment still hinders the applicability of these techniques in many scenarios. Here, we overcome this challenge and demonstrate the manipulation of objects in disordered and dynamic media, by optimally tailoring the momentum of sound waves iteratively in the far field. The method does not require information about the object's physical properties or the spatial structure of the surrounding medium but relies only on a real-time scattering matrix measurement and a positional guidestar. Our experiment demonstrates the possibility of optimally moving and rotating objects, extending the reach of wave-based object manipulation to complex and dynamic scattering media. We envision new opportunities for biomedical applications, sensing, or manufacturing.
\end{abstract}

\keywords{Object manipulation, wave-momentum shaping, scattering, adaptive acoustics.}
\maketitle
%

Ever since the emergence of optical tweezers \cite{Ashkin:86,ashkin_optical_1987}, the non-contact manipulation of objects using electromagnetic \cite{ramachandran_review_2022,bustamante_optical_2021} and acoustic waves \cite{hagsater_acoustic_2007,hagsater_acoustic_2008,andrade_acoustic_2020} has become a central paradigm in quite diverse fields ranging from optomechanics to bio-acoustics. Sound waves, in particular, offer distinct advantages, as they are bio-compatible and harmless, while their short wavelengths can penetrate a wide range of heterogeneous, opaque, and absorbing media. Another key feature of acoustics is its wide frequency range, spanning from Hertz to Gigahertz, which facilitates the manipulation of particles varying in size from a few centimeters to a few micrometers. In this way, not only Mie  \cite{marzo_acoustic_2018,andrade_acoustic_2017,zhao_standing_2011} and Rayleigh particles can be addressed, but also complex objects including individual biological cells \cite{Ozcelik_Rufo_Guo_Gu_Li_Lata_Huang_2018,rufo_acoustofluidics_2022,dholakia_comparing_2020}. 

While various strategies have already been developed to collectively or selectively manipulate objects and particles, these techniques always rely on controlled and static environments.  Collective dynamic positioning of particles trapped in the potential wells of a pressure field has been achieved in 1D \cite{whymark_acoustic_1975}, 2D \cite{foresti_acoustophoretic_2014, collins_two-dimensional_2015} or 3D \cite{hoshi_three-dimensional_2014,marzo_holographic_2015,prisbrey_ultrasound_2018,franklin_three-dimensional_2017}. Typically, by generating appropriate standing waves \cite{zang_standing_2020}, particles or objects are trapped either on the pressure nodes or antinodes, depending on their contrast ratio with the surrounding fluid \cite{dholakia_comparing_2020}. More advanced strategies have also been developed to address the selectivity problem of standing-wave-based trapping, involving acoustic vortices \cite{baresch_observation_2016}, or the use of additional systems such as lenses \cite{wang_particle_2016}, metasurfaces \cite{memoli_metamaterial_2017} or holograms \cite{marzo_holographic_2015,marzo_holographic_2019,baudoin_folding_2019,baudoin_spatially_2020,melde_holograms_2016}. Considerable attention has also been paid to the development of on-chip acoustofluidic and acoustophoretic devices  \cite{rufo_acoustofluidics_2022,nguyen_acoustofluidics_2023, fan_recent_2022, yang_harmonic_2022} and wave-controlled micro-robots~\cite{fischer_tiny_2021,lee_needle-type_2020,durrer_robot-assisted_2022,zhou_magnetically_2021,bachelard2017emergence,zhang_rolling_2022,peters_superparamagnetic_2014} for lab-on-a-chip and biomedical applications. 
However, the requirement for precisely controlled static environments and proximity to the target significantly restricts the applicability of these various techniques in many real-world scenarios. Practical cases involve disordered or dynamic environments where manipulation must occur at a considerable distance from the object that needs to be manipulated.

Here, we propose and experimentally demonstrate a wave momentum shaping approach, which only requires far-field information and allows us to move and rotate objects even in disordered or dynamic environments. Instead of relying on potential wells to trap the object, we continuously find and send the optimal mode mixture that transfers an optimal amount of momentum to the object. This mode mixture is updated during the motion as the scattering changes. The method is experimentally demonstrated in a macroscopic 2D acoustic cavity containing a movable object and a collection of scatterers. Far-field scattering matrix measurements allow us to determine the optimal wavefronts for shifting or rotating the object at each moment in time. Remarkably, the method neither requires the knowledge or modelling of acoustic forces nor any prior information on the physical properties of the object or disorder. Only a guidestar measurement of the object's position or rotation angle is needed, which is here provided by a camera. The remarkable robustness of the method is emphasized by implementing it in a dynamic scenario, where the scatterers composing the environment move randomly. The method may be transposed to other platforms and scales, such as ultrasounds or light for the motion of microscopic bodies.

\begin{figure*}%
\centering
\includegraphics[width=\textwidth]{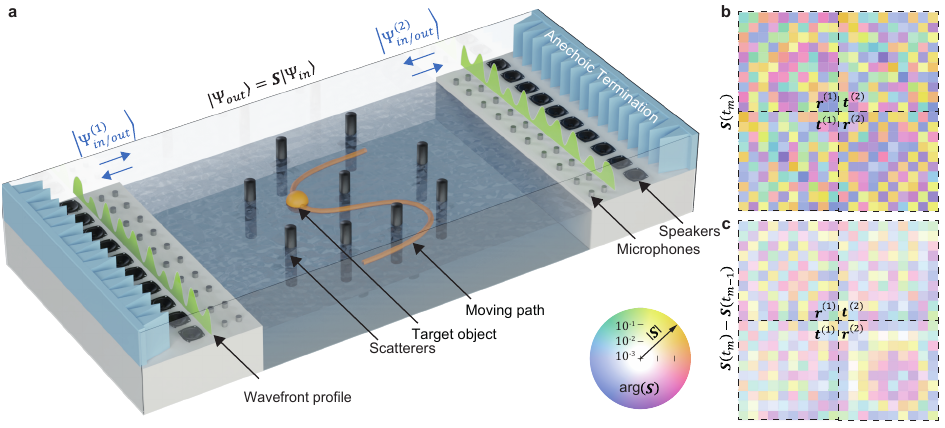}
\caption{\textbf{Moving an object in a complex scattering medium by acoustic wave-momentum shaping.} \textbf{a}, We consider a parallel plate acoustic waveguide supporting 10 modes at the working frequency and containing cylindrical rigid scatterers (in black). The bottom surface of this waveguide is formed by the water in this container, allowing a spherical object to float and move freely (orange ball). Wave momentum shaping consists of finding and sending, at each time $t_m$, the optimal mode mixture to push the ball along an arbitrarily chosen path (orange line). We achieve this by real-time far-field measurements, allowing us to track the evolution of the scattering matrix $\boldsymbol{S}$ as the object moves, deducing the wavefronts to be injected by the external speaker arrays to optimally deliver the target momentum. \textbf{b}, Example of a scattering matrix measured at a given time $t_{m}$ in our experiment. \textbf{c}, Difference between the scattering matrix at $t_{m}$ and the one measured at $t_{m-1}$, showing the influence of a small object translation on scattering. We use the information collected at three consecutive time steps to derive the mode mixture that optimally pushes the ball in the desired direction. The static scatterers are later replaced with dynamic ones.}\label{fig1}
\end{figure*}

\subsection*{Principles of wave-momentum shaping} \label{main}
The idea of wave-momentum shaping is inspired by recent developments in adaptive optics and disordered photonics, where wavefront shaping techniques have been significantly advanced to focus light in disordered media or to compensate aberrations and multiple-scattering for various purposes \cite{vellekoop_focusing_2007,popoff_measuring_2010,gigan_roadmap_2022, yu_wavefront_2022, cao_shaping_2022}. In the most straightforward implementation, a feedback mechanism allows a quantity of interest, such as the optical power focused at a given point, to be iteratively optimized by tuning the incident wavefronts \cite{vellekoop_focusing_2007}. On the other hand, more advanced concepts, such as Wigner-Smith operators derived from a system's scattering matrix, have provided ways to focus light in disorder optimally \cite{ambichl_focusing_2017}, exert a maximal electromagnetic force or torque on static objects \cite{horodynski_optimal_2020,horodynsky_tractor_2023}, or potentially even cool an ensemble of levitated particles  \cite{hupfl2023optimalb,hupfl_optimal_2023}. Wave-momentum shaping applies these ideas to the manipulation of moving objects, combining the optimal character of Wigner-Smith approaches with iterative guidestar techniques, necessitated by the dependence of the $\boldsymbol{S}$ matrix on the object position, which influences the complex scattering process, constantly modifying the field speckle.

Consider the experimental setup illustrated in Fig.~\ref{fig1}\textbf{a}, consisting of an acoustic multimode waveguide supporting ten modes ($N=10$) at the operational frequency $f_0$ = 1590 Hz (audible sound). In the central part, we introduced a movable object (ping-pong ball) with a radius of 20 mm ($\approx 0.1 \lambda_0$), which floats on the surface of a water tank. Within this tank, multiple static cylindrical scatterers (depicted as black cylinders) protrude above the water level, thus creating a complex scattering landscape. Two arrays of ten speakers are placed on both sides, labeled 1 and 2, allowing us to control the incident acoustic mode mixtures $\mid\boldsymbol{\Psi}_{in}^{(1,2)}\rangle$. These incident waves are linearly scattered into outgoing mode mixtures $\mid\boldsymbol{\Psi}_{out}^{(1,2)}\rangle$, which can be measured using microphones placed in the waveguide's asymptotic regions (see Methods and Supplementary Information). From such measurements, it is possible to deduce the scattering matrix $\boldsymbol{S}(t)$, which evolves with time because, as the target object moves, it modifies the scattering occurring in the central region. This dynamic scattering matrix obeys the relation $\mid\boldsymbol{\Psi}_{out}(t)\rangle=\boldsymbol{S}(t)\mid\boldsymbol{\Psi}_{in}(t)\rangle$, where we gathered the states related to both sides into single column vectors.

\begin{figure*}[htpb]%
\centering
\includegraphics[width=0.7\textwidth]{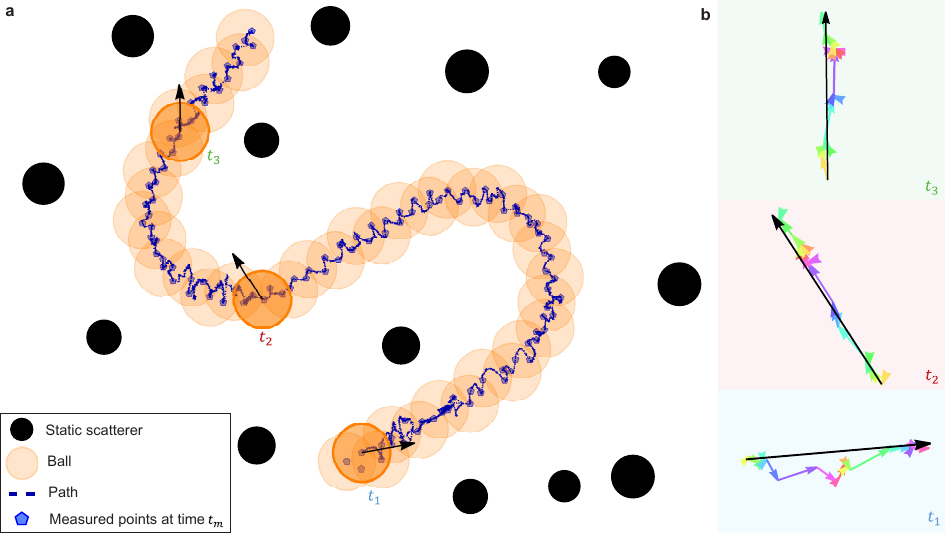}
\caption{\textbf{Experimental demonstration of object guiding through acoustic wave-momentum shaping in a static scattering medium}. \textbf{a}, A set of points, in blue, are chosen to define an overall S-shaped path to be followed by the moving ball, whose successive positions captured by a camera are shown by orange disks. The ball successfully reaches each blue point, where the $\boldsymbol{S}$ matrix is measured. Crucially, these checkpoints are chosen to zigzag about the S-shaped path, so that the last three consecutive measurements contain optimal information on the gradient of the $\boldsymbol{S}$ matrix with respect to the ball coordinates. \textbf{b}, Net momentum imparted to the ball at three different times on the path (black arrow), and its decomposition over the modes injected from the two sides. Note that these are not derived from the ball dynamics, but rather inferred from the momentum expectation value of the injected Wigner-Smith eigenstates. Remarkable agreement with the actual direction of the ball velocity, reported in panel \textbf{a} (black arrows), is observed. See supplementary movie SM1.}
\label{fig2}
\end{figure*}

Figure~\ref{fig1}\textbf{b} shows an example of the measured scattering matrix, of dimension $2N\times2N$, which is composed of four $N\times N$ sub-blocks, the reflection and transmission matrices $\boldsymbol{r}^{(1),(2)}$ and $\boldsymbol{t}^{(1),(2)}$, describing how each of the ten modes scatter on each side. The figure encodes the amplitude of the matrix coefficients in the transparency of the squares, and the phase in their color. Clearly, mode mixing occurs due to the presence of complex scattering. Quite intuitively, to make object manipulation possible, such a scattering matrix must depend on the position of the movable object. This dependence is evidenced by repeating the scattering matrix measurement after slightly moving the object (by a distance equal to a quarter of its radius, i.e., $l=5$ mm), and plotting the difference in Fig.~\ref{fig1}\textbf{c}. We observe that while the changes are small in magnitude, consistent with the fact that only one scatterer is moved, some information about the object motion seem to be embedded in scattering phase changes.

The $\boldsymbol{S}$ matrix's dependence on the object's position is relevant in the context of its dynamic manipulation. If we denote by $\alpha$ either the $x$ or $y$ coordinate of the movable object, the momentum transferred to it upon scattering $\Delta p_{\alpha}$ can be calculated via the expectation values of the operator $\boldsymbol{C}_{\alpha} = -\textrm{i}\partial/\partial \alpha$ for the superposition states  $\mid\boldsymbol{\Psi}_{in,out}\rangle$ \cite{horodynski_optimal_2020}. The momentum transferred onto the particle upon scattering is the difference between the momentum of the outgoing and incident mode mixtures in the vicinity of the particle. Assuming unitary scattering ($\boldsymbol{S}^{\dag}\boldsymbol{S}=\mathbb{1}$), one can demonstrate the link between this momentum transfer and variation of $\boldsymbol{S}$ with $\alpha$ (Methods):

\begin{equation}
\begin{aligned}
    \Delta p_{\alpha}=\langle\boldsymbol{\Psi}_{in}\mid \left(-\textrm{i}\boldsymbol{S}^{-1}\frac{d\boldsymbol{S}}{d\alpha}\right) \mid\boldsymbol{\Psi}_{in}\rangle.
\end{aligned}
\label{eq:WSO}
\end{equation}
The Hermitian operator $\boldsymbol{Q_\alpha}=-\textrm{i}\boldsymbol{S}^{-1}d\boldsymbol{S}/d\alpha$ is known as a generalized Wigner-Smith operator \cite{ambichl_focusing_2017}. Equation \eqref{eq:WSO} means that the momentum imparted locally onto the moving object upon scattering is related to the expectation value of $\boldsymbol{Q_\alpha}$ for the specific input state $\mid\boldsymbol{\Psi}_{in}\rangle$ in the far field. A direct consequence of Eq.~\eqref{eq:WSO} is that if the input state $\mid\boldsymbol{\Psi}_{in}\rangle$ is chosen to be an eigenvector of $\boldsymbol{Q_\alpha}$, the momentum kick on the object will be proportional to its eigenvalue. Therefore, choosing the eigenstate with the highest eigenvalue as the input mode mixture will optimize the transfer of momentum to the object in the direction $\alpha$. This is the basic physical principle behind wave-momentum shaping.

\subsection*{Linear-momentum transfer}
\begin{figure*}[htpb]%
\centering
\includegraphics[width=0.7\textwidth]{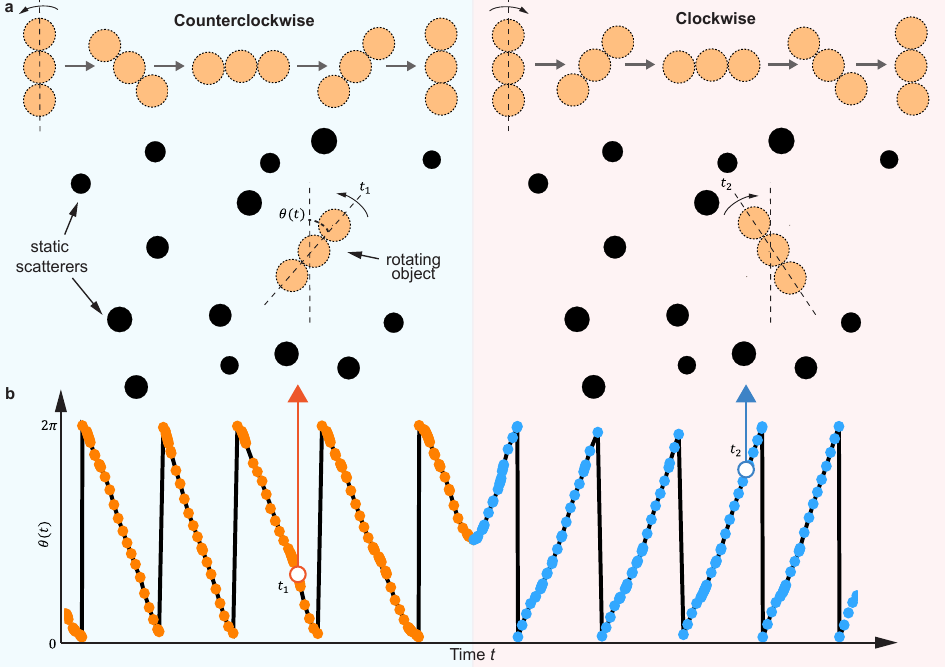}
\caption{\textbf{Experimental demonstration of object rotation by acoustic angular-momentum shaping in static scattering media}. \textbf{a}, We use audible sound to rotate an object constructed from three balls glued together in a disordered medium. First, we move the target in the counter-clockwise direction (left part of the figure), and then abruptly switch its direction of rotation (right part). At each step (10 degrees), we extract from a far-field $\boldsymbol{S}$ matrix measurement the Wigner-Smith operator with respect to the rotation angle $\theta$, which allows us to send the wavefront with maximal angular-momentum transfer. \textbf{b}, Measured angle versus time, confirming the rotation of the object, in the anti-clockwise then clockwise directions. }
\label{fig3}
\end{figure*}

We first apply wave-momentum shaping to the transfer of linear momentum, and experimentally demonstrate complete control over the trajectory of a moving object in a complex scattering medium, which is static for now. We start from the set-up of Fig.~1 and apply an iterative motion algorithm that works as follows: (i) Initially, the object is at rest. We send three random wave fields to move it slightly but randomly, and measure the $\boldsymbol{S}$-matrix at three different nearby points, whose positions are measured by the camera; (ii) From these measurements, we estimate the components $d\boldsymbol{S}/d\alpha$ of the gradient of $\boldsymbol{S}$ with respect to the coordinates $\alpha=x,y$, using discrete derivative approximations; (iii) We compose $\boldsymbol{Q_\alpha}$ and diagonalize it to obtain mode mixtures and momentum expectations (eigenvalues) in $\alpha=x,y$; (iv) We send a superposition of eigenvectors of $\boldsymbol{Q_x}$ and $\boldsymbol{Q_y}$ in proportion, to move the object in a desired direction and measure $\boldsymbol{S}$ again once the object has moved; (v) the process is iteratively repeated based on the last three measured $\boldsymbol{S}$ matrices until the object arrives at the desired destination. The method does not require calibration or access to the interior of the medium.

Figure~\ref{fig2}\textbf{a} demonstrates the successful guiding of an object within a disordered medium using acoustic wave-momentum shaping. Several snapshots of the moving ball are blended into one picture to illustrate better the path followed by the moving scatterer. A video recorded by our camera can be found in Supplementary movie SM1. Remarkably, the acoustic fields injected from the far field are able to continuously move the floating ball through a chosen S-shaped trajectory within the disordered medium (the total path length is around four $\lambda_0$). It is worth noting that the path is discretized into intermediate checkpoints (blue disks) arranged in a zig-zag manner about the S-shaped trajectory to enable a good estimation of the $\boldsymbol{S}$ matrix gradient (see Section 1.6 and Fig.~S6 of S.I.). Note that the object is not trapped, but moved by successive acoustic pushes much like a hockey player guiding a puck.

To illustrate the contribution of each mode and in which sense the input state at each time step is optimal, Fig.~\ref{fig2}\textbf{b} compares at three distinct times the momentum expectation value of the input superposition (black arrow) with the ones of its individual modes components alone (colored arrows). It is clear that each mode contributes to pushing the ball in the correct final direction, and the total push is due to a collective action of all modes. We note that some modes do not push the ball exactly in the desired direction. Yet, this mixture is optimal given the constraints on the wave spatial degrees of freedom imposed by the disordered medium at this specific location. We also compare the total momentum expectation (black arrows in Fig.~\ref{fig2}\textbf{b}), which is a theoretical prediction, to the actual velocity of the ball (black arrows in Fig.~\ref{fig2}\textbf{a}), which is an experimental observation. The remarkable agreement between the directions of the expected momentum push, and those of the measured velocity confirms that we successfully implemented our wave-momentum shaping strategy. We conclude that the unavoidable absorption losses present in any experiment, which alter the fields' amplitudes more than their phases, does not significantly influence the direction of the momentum push predicted by the unitary theory. The interested reader will find other path instances in Supplementary movie SM2.

\begin{figure*}[htpb]%
\centering
\includegraphics[width=0.7\textwidth]{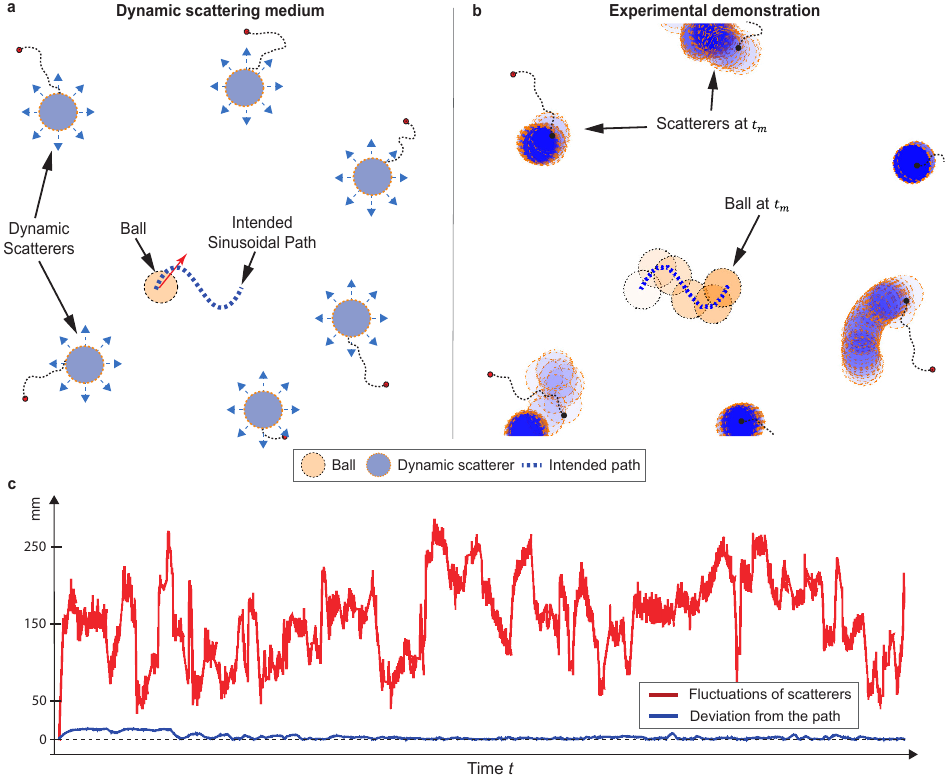}
\caption{\textbf{Moving a specific object among a dynamic ensemble of scatterers experiencing random motion}. \textbf{a}, We let all scatterers move freely under fast external perturbations, and wish to control the trajectory of the orange ball, guiding it on a sinusoidal path. The blue balls have a metallic nut glued to them, allowing us to randomize their motion by applying fast magnetic perturbations with a moving external electromagnet. They are loosely anchored to the ground by strings to avoid any collision with the orange ball. \textbf{b}, Experimental trajectories measured by a camera, demonstrating the successful control of the ball trajectory even in this extreme dynamic scenario. \textbf{c}, Comparison between the measured deviation of the target ball center from the intended sinusoidal path (blue), and the large fluctuations of the other scatterers from their initial positions (red).}
\label{fig4}
\end{figure*}

\subsection*{Angular-momentum transfer}
An advantage of the variational principle presented above is that $\alpha$ is not restricted to be the $x$ or $y$ coordinate, but can be any observable target parameter influencing the scattering. A relevant example we consider in the following is the rotation angle $\theta$ of an object. This choice will allow us to create an acoustic motor and rotate objects from a distance, by sending audible sound. Consider the angular-momentum transfer on a rotating object constructed from three balls glued together, and placed on a fixed rotation axis at its center, located within the disordered medium (Fig. \ref{fig3}\textbf{a}). The instantaneous scattering matrix $\boldsymbol{S}(t_m)$ is here measured at consecutive time instances $t_m$ with 20 degrees angle step, harnessing the angular momentum operator $\boldsymbol{Q_\theta}$, and providing a way to induce optimal transfer of torque from the field to the object. Figure~\ref{fig3}\textbf{b} reports the experimentally measured value of $\theta$ as a function of time. In this experiment, we first selected eigenvectors of $\boldsymbol{Q_\theta}$ with positive eigenvalues, consistent with the counter-clockwise rotation initially observed during the experiment (blue-shaded part of the figure). Then, we abruptly switched to input states with negative eigenvalues (red shaded part). The observation of a reversal of the rotation direction, reported in Fig.~\ref{fig3}\textbf{b}, is thus consistent with theoretical expectations (Supplementary Movie SM3). 

\begin{figure*}[htpb]%
\centering
\includegraphics[width=0.7\textwidth]{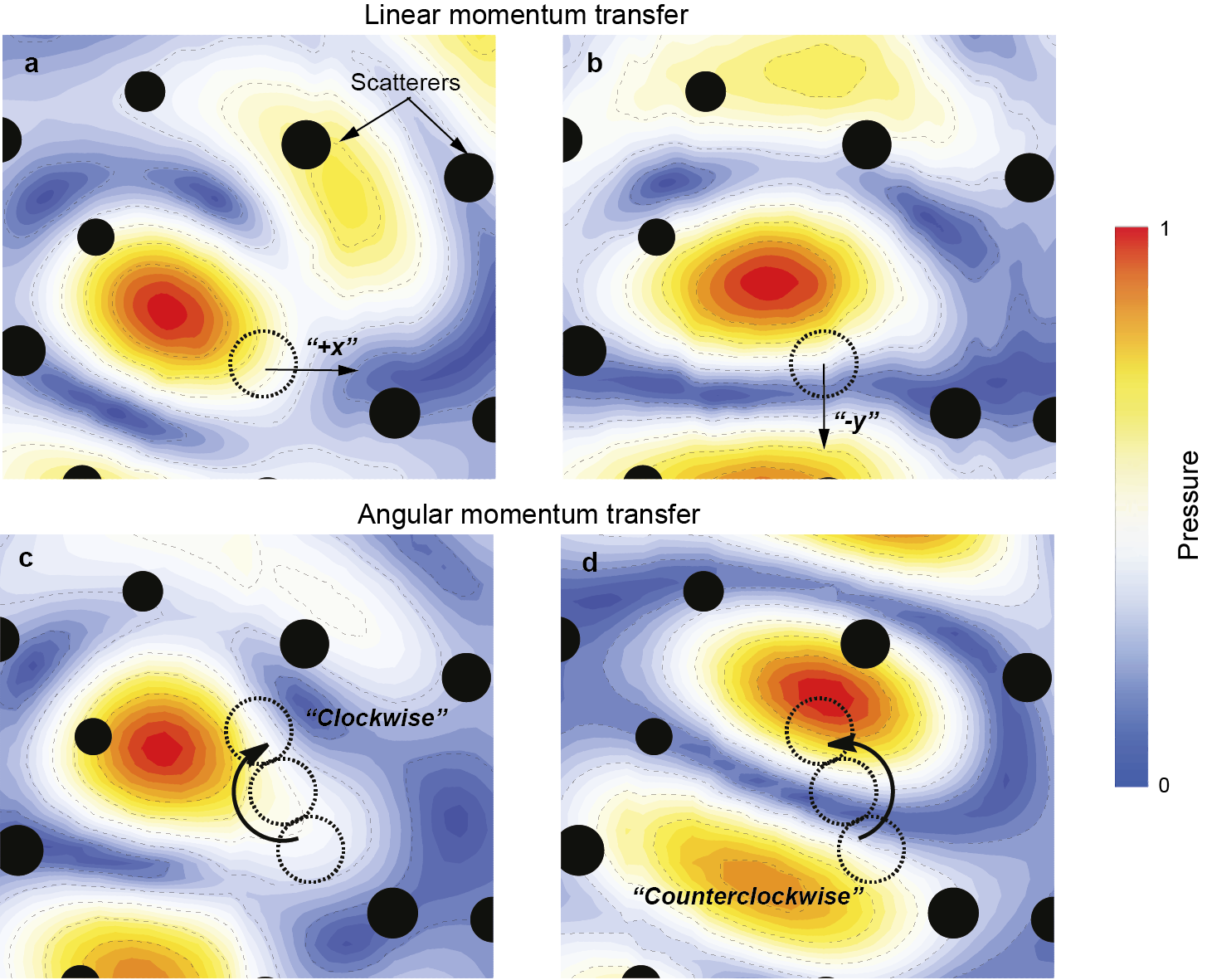}
\caption{\textbf{Experimental field scans around the moving object}. We measured the distribution of acoustic pressure amplitudes associated with the optimal transfer of linear momentum along the $x$ and $-y$ directions (panels \textbf{a} and \textbf{b}, respectively). The case of the rotating object is illustrated in \textbf{c} and \textbf{d} for clockwise and counterclockwise rotations, respectively. The cylindrical scatterers are represented by black disks, and the moving object is outlined with dashed circles. With our approach, we can create the optimal speckle field allowed by the scattering medium, automatically creating the best possible hot spot next to the object in order to achieve each prescribed momentum kick.}
\label{fig5}
\end{figure*}

\subsection*{Manipulating in dynamic disorder}
Since the manipulation method is based on real-time measurements of the instantaneous scattering matrix, nothing prevents the scattering environment from changing in time as well. To demonstrate this, we added other floating balls (similar to the moving target ball) inside the cavity. This experiment and its goal are illustrated in Fig.~\ref{fig4}\textbf{a}. The ball we want to control is the orange one, while the blue ones are the added balls, which are anchored with light strings to prevent any collision with the target ball. The blue balls carry small metallic nuts, which allows us to randomize their motion by varying in time the magnetic field inside the cavity. We wish to control the path of the orange ball and make it follow a shape that looks like a period of a sine function (dashed blue line). Panel \textbf{b} shows the measured successive positions of all balls during the experiment. Contrary to the blue balls, which move unpredictably, the orange ball closely follows the pre-designed sinusoidal path. The deviations of the orange ball from this target path, shown as a blue line in \textbf{c}, are tiny. For comparison, we also plot as a red line the average distance browsed by the blue balls away from their initial position, which fluctuates more strongly in magnitude and speed, underlining the extreme control that we can maintain on the trajectory of the target. A video showing the robustness of the manipulation in the dynamically changing random medium in Fig.~\ref{fig2}\textbf{c} is provided as Supplementary Movie SM4.

\subsection*{Acoustic pressure field maps}
We provide another point of view on wave-momentum shaping by probing the acoustic pressure field in the vicinity of the moving object (Methods). For this purpose, we exert the optimal linear momentum push of a ball along different directions, such as $+x$, or $-y$, and rotations in opposite directions, and measure the acoustic field map. From the displayed pressure profiles (Fig.~\ref{fig5}\textbf{a}-\textbf{d}), we see that the speckle field tends to create hot spots of acoustic pressure to push the object in the right direction. Conversely, experimental pressure distributions for the eigenvalues with the smallest absolute value of the corresponding Wigner-Smith operator exhibit no hot spot near the particle and tend to put it in a silent zone (Supplementary Fig. S8). Note that input states that are optimal for motion along $x$ can still exhibit a non-zero expectation value in the $y$ direction. However, they remain the most efficient at pushing along $x$. Therefore, combining eigenvectors to control expectation values in unwanted directions may provide a way to further refine the algorithm and the precision of the motion. Sometimes the object finds itself in a location in the medium where it can't be pushed in the right direction given the available degrees of freedom in the speckle. This is not a problem since these points are isolated, and the algorithm will make the object catch up with the trajectory at the next points. To conclude, it is striking to observe how the optimal pressure field can be prepared around the object without knowing anything about the involved wave-matter interaction, nor about the object's shape or environment. This is a clear advantage of the present method when compared to conventional methods based on trapping.

\section*{Conclusion}
In this work, we report the experimental control of an object's translation and rotation in a complex and dynamic scattering medium through the concept of wave-momentum shaping. An iterative manipulation protocol, based solely on the knowledge of the far-field scattering matrix of a system and a position guidestar, enables the optimal transfer of linear and angular momentum from an acoustic field for object manipulation within both static and dynamic disordered media. The dynamically injected wavefronts generate the optimal field speckle near the object to be moved, much like a hockey player guiding a puck, in order to produce successive momentum kicks. This method is free of potential traps, robust against disorder, and tolerates that the surrounding medium changes in time throughout the manipulation. Remarkably, the method is rooted in momentum conservation and does not require any knowledge of the object to be manipulated, but only a guidestar measurement of its position. In addition, it does not require any modelling of interaction forces, making the protocol very general and broadly applicable to many real-life scenarios (including different waves, scales, objects, etc.). Future efforts will focus on developing methods for objects of various sizes, for example by transposing the concept to ultrasonic frequencies for the manipulation of smaller objects, as well as extensions to the control of multiple objects. For this, we note that the frequency degree of freedom could also be leveraged. We conclude that the method seems particularly promising for micromanipulation and acoustofluidic applications such as tissue engineering \cite{olofsson_ultrasonic_2018,armstrong_engineering_2018}, biological analysis \cite{rufo_acoustofluidics_2022}, and drug delivery \cite{baresch_acoustic_2020,xu_acoustic_2023}, among others.

\section*{Methods}\label{methods}

\subsection*{Experimental setup} 
The setup consists of a water-filled tank (Figs.~\ref{fig1} \textbf{a} and S1 \textbf{a}) coupled at the top to a two-dimensional air waveguide terminated at both ends by anechoic terminations. The water tank's width, length, and height are $100\times100\times3$ cm, respectively. The 2D acoustic waveguide above it has a width of 104 cm, a length of 180 cm, and a height of 8 cm. Two columns of 10 ICP{\textregistered} microphones (PCB 130F20, 1/4 inch) separated by 5 cm are placed between the tank and the anechoic terminations on each side to measure the complex pressure field distribution inside the waveguide. Two columns of 10 amplified loudspeakers (Monacor MSH-115, 4 inches, with in-house amplifiers) are placed horizontally on each side's bottom of the air waveguide to ensure the efficient excitation of all ten modes inside the acoustic structure. The generation of incident wave states and the acquisition of the corresponding far-field scattering are made using an FPGA Speedgoat Performance Real-Time Target Machine controller (I/O 135, sampling rate 10 kHz) with 40 inputs and 20 outputs. The controller is programmed by Matlab/Simulink{\textregistered} to generate the proper acoustic wavefronts by controlling the voltage of 20 loudspeakers and to acquire voltages produced by 40 microphones corresponding to the pressure signals. PCB Piezotronics 483C05 sensor conditioners are used to pre-condition the microphone signals. Examples of captured signals and post-processing are shown in Fig.~S2. 

The moving target scatterer is a ping-pong ball (diameter 4 cm and weight 4.17 g) floating on the water's surface. The static disorder scatterers are plastic cylinders of various diameters (2 to 4 cm) immersed in the water tank and partially surfaced without reaching the upper part of the 2D air waveguide. The cylinders and the ball are waxed (coated with candle wax, \textit{i.e.}, a hydrophobic film) to prevent the ball from sticking to the static scatterers by capillarity. The real-time position of the ball (Fig.~S4) is captured by an ultra-wide Logitech Brio webcam, working in Full HD resolution ($1920 \times 1080 pixels$) and high refresh rate (60 frames per second) mode. The moving target is placed at the initial (starting) position with the help of a small iron nut glued to it and the electromagnet attached to the mechanical arm (Fig.~S1 \textbf{b}), which is moved in a volume ($1000 \times 1000 \times 110$ mm) above the water tank by three high-precision linear stages (Newport{\textregistered} IMS stages with displacement error < 0.05 mm).

The rotating object of Fig.~\ref{fig3} comprises 3 ping-pong balls glued together in a line and placed on a fixed needle in its center to prevent linear displacement and only allow rotation while limiting friction. The balls forming it are painted with different patterns to facilitate the detection of the instant angle value.

To create the dynamic scattering medium, we used ten ping-pong balls and glued small metallic nuts on them. The scattering balls, evenly positioned around the intended paths, are attached to the bottom of the water tank by 3- to 8-cm-long nylon threads. The random fluctuations of these scatterers are then exacerbated by randomly moving the mechanical arm over the balls while randomly switching the state of the attached electromagnet. The disorder scatterers are placed at a significant enough distance from the target scatterer, which is still free-floating, to avoid any collision.

Finally, the top plates above the water tank are replaceable with carefully designed perforated plates (holes with a diameter of 1 mm and forming a square array with a period of 10 mm, Fig.~S1 \textbf{e} to allow scanning the field inside the waveguide from a microphone placed on the robotic arm, which is located outside the waveguide.

\subsection*{Scattering matrix measurement}
The complex scattering matrix $\boldsymbol{S}$ relates the incoming with the outgoing flux-normalized modes through a set of $2N$ linearly independent equations ($2N=20$ is the total number of propagative modes, 10 from each side): $\boldsymbol{\Psi}_{out} = \boldsymbol{S} \cdot\boldsymbol{\Psi}_{in}$. Solving the scattering matrix $\boldsymbol{S}$ requires measuring $N$ independent wave mode distributions excited by a combination of speakers that form an orthogonal basis. Our experiment uses an orthonormal basis where only one speaker is excited at a time with a 1590 Hz harmonic signal. For each excitation, the data collected by the microphone arrays on both sides can be used to determine the incident and outgoing modes $\boldsymbol{\Psi}_{in, out}$. With the hardware we used, this takes about $80$ ms. Therefore, after $2N$ orthogonal excitations ($1.6$ s), the scattering matrix is solved for that particular scattering configuration. Such raw scattering matrix is neither perfectly symmetric, nor unitary, and is subsequently regularized, by discarding its very small antisymmetric part and rescaling its subunitary eigenvalues, keeping their phases (see Fig.~S3). 

\subsection*{Construction of Generalized Wigner-Smith Operators}
The construction of the GWS operator $Q_\alpha$ is based on gradient approximations, which require successive measurements of the scattering matrix $\boldsymbol{S}(t)$ at three different positions (time).

To derive the translation GWS operators $\boldsymbol{Q}_{x}$ and $\boldsymbol{Q}_{y}$ for the ball at position ($x_m,y_m$) and time instance $t_m$,  we need, in addition to the scattering matrix $\boldsymbol{S}_m$ measured at the actual position, the scattering matrices $\boldsymbol{S}_{m-1}$ and $\boldsymbol{S}_{m-2}$ measured at the two previous time-instances $t_{m-1}$, $t_{m-2}$, when the ball was located at coordinates $(x_{m-1},y_{m-1})$ and $(x_{m-2},y_{m-2})$ respectively.

With these three matrices $\boldsymbol{S}_m$,  $\boldsymbol{S}_{m-1}$, and $\boldsymbol{S}_{m-2}$, the gradient of $\mathbf{\boldsymbol{S}}$ can be derived with the following approximation formulae
\begin{equation}
    \begin{bmatrix}
      \partial_x \mathbf{\boldsymbol{S}}_m \\
      \partial_y \mathbf{\boldsymbol{S}}_m 
    \end{bmatrix} \approx
    \begin{bmatrix} x_m - x_{m-2} & y_m - y_{m-2} \\
        x_m - x_{m-1} & y_m - y_{m-1} 
    \end{bmatrix}^{-1}\cdot
    \begin{bmatrix} 
        \boldsymbol{S}_m-\boldsymbol{S}_{m-2} \\\boldsymbol{S}_{m}-\boldsymbol{S}_{m-1} 
    \end{bmatrix}   . 
    \label{eq:gradWSO}
\end{equation}

With the gradient estimated, the construction of the GWS operators $\boldsymbol{Q}_{x},\boldsymbol{Q}_{y}$ is direct and reads as

\begin{align*}
    \boldsymbol{Q}_{x} &=-\textrm{i}\boldsymbol{S}_m^{-1}\partial_x\boldsymbol{S}_m,\\
    \boldsymbol{Q}_{y} &=-\textrm{i}\boldsymbol{S}_m^{-1}\partial_y\boldsymbol{S}_m.
\end{align*}
 The error in the gradient approximation and, therefore, in the operators $\boldsymbol{Q}_{x}$ and $\boldsymbol{Q}_{y}$ depends on the shape of the triangle formed by the three measurement points, with the best results obtained for an equilateral triangle and worse for a flat scalene one (see Fig.~S6). Detailed analysis of the triangle's influence on the derivation of GWS is provided in Supplementary Information. Therefore, for the best manipulation of the object position, the moving path is drawn with a zig-zag line to minimize the error in the GWS operators. 

Similarly, the rotation GWS operator $\boldsymbol{Q}_{\theta}$ requires the measurement of $\boldsymbol{S}_m$,  $\boldsymbol{S}_{m-1}$, and $\boldsymbol{S}_{m-2}$, for three consecutive vane angles $\theta_m$, $\theta_{m-1}$, and $\theta_{m-2}$, taken at time $t_{m}$, $t_{m-1}$, and $t_{m-2}$.

The gradient approximation is in that case derived with a backward three-point derivative

\begin{equation}
    \partial_{\theta}\boldsymbol{S}_{m} \approx \frac{\boldsymbol{S}_{m}-4\boldsymbol{S}_{m-1}+3\boldsymbol{S}_{m-2}}{\delta \theta},
\end{equation} 
where $\delta \theta$ is the angle difference between the time instance $t_{m}$ and $t_{m-1}$. 

The GWS operator constructed to control the rotation of the vanes therefore reads as follows

\begin{equation}
    \boldsymbol{Q}_{\theta} =-\textrm{i}\boldsymbol{S}_m^{-1}\partial_\theta\boldsymbol{S}_m.
\end{equation}


\subsection*{Injection of optimal input mode mixtures}
As explained in the main text, finding the optimal mode mixture to be injected to give the optimal momentum push to the object follows from Eq. (1). We, therefore, provide a short proof for this important equation.

For a particle in free space, the change of momentum transferred to it upon scattering $\Delta p_{\alpha}$ can be calculated via the expectation values of the operator $\boldsymbol{C}_{\alpha} = -\textrm{i}\partial/\partial \alpha$ for the superposition states  $\mid\boldsymbol{\Psi}_{in,out}\rangle$

\begin{equation}
\begin{aligned}
    \Delta p_{\alpha}=<\boldsymbol{\Psi}_{out}\mid\boldsymbol{C}_{\alpha}\mid\boldsymbol{\Psi}_{out}>-<\boldsymbol{\Psi}_{in}\mid\boldsymbol{C}_{\alpha}\mid\boldsymbol{\Psi}_{in}>,
\end{aligned}
\end{equation}

We can then write

\begin{equation}
\begin{aligned}
    \Delta p_{\alpha}=-i\left(\boldsymbol{\Psi}_{out}\mid\frac{d}{d\alpha}\mid\boldsymbol{\Psi}_{out}>-<\boldsymbol{\Psi}_{in}\mid\frac{d}{d\alpha}\mid\boldsymbol{\Psi}_{in}>\right).
\end{aligned}
\end{equation}

Using $\boldsymbol{S}^{\dag}\boldsymbol{S}=1$, we can write $<\boldsymbol{\Psi}_{in}\mid d/d\alpha \mid\boldsymbol{\Psi}_{in}>=<\boldsymbol{S \Psi}_{in}\mid \boldsymbol{S} d/d\alpha \mid\boldsymbol{\Psi}_{in}>$ and obtain 

\begin{equation}
\begin{aligned}
    \Delta p_{\alpha}=-i<\boldsymbol{S \Psi}_{in}\mid \left(\frac{d}{d\alpha}\mid\boldsymbol{S \Psi}_{in}>-\boldsymbol{S}\frac{d}{d\alpha}\mid\boldsymbol{\Psi}_{in}>\right).
\end{aligned}
\end{equation}

The term in parentheses is nothing but $\frac{d \boldsymbol{S}}{d\alpha}\mid\boldsymbol{\Psi}_{in}>$, and we directly get

\begin{equation}
    \Delta p_{\alpha}=-i<\mathbf{\boldsymbol{S \Psi}_{in}}\mid \frac{d \boldsymbol{S}}{d\alpha}\mid\boldsymbol{\Psi}_{in}>=<\boldsymbol{\Psi}_{in}\mid -i\boldsymbol{S}^{\dag}\frac{d\boldsymbol{S}}{d\alpha}\mid\boldsymbol{\Psi}_{in}>. 
\end{equation}
In Refs. \cite{ambichl_focusing_2017, horodynski_optimal_2020}, it was shown that this relation, which is Eq. (1) in the main text, continues to hold even when the target particle is embedded in a scattering environment. In this way, the momentum push expected for a given far-field input is expressed as the expectation value of the generalized Wigner-Smith operator $\boldsymbol{Q_\alpha}$, which is Hermitian. Therefore, the optimal momentum push is provided by the eigenvector of $\boldsymbol{Q_\alpha}$ with the highest eigenvalue. 

Having measured the Wigner-Smith operators, we diagonalize them and find the eigenvector with the highest eigenvalue, and use them to calculate the optimal mode mixture to be injected to give the optimal momentum push to the particle. For example, if we want to move the object by $\Delta x$ and $\Delta y$: (i) we diagonalize $\boldsymbol{Q}_{x}$ and $\boldsymbol{Q}_{y}$; (ii) we obtain their eigenvectors with highest eigenvalues, $\Psi_{x,y}$, with eigenvalues calculated as $\delta x$ and $\delta x$; (iii) we construct the optimal input state $\frac{\Delta x}{\delta x}\Psi_{x}+\frac{\Delta x}{\delta x}\Psi_{y}$. This input state is multiplied with the coupling coefficients matrix of the speakers \textbf{M}, obtaining the required voltage amplitudes and phases required on each speaker (see Fig.~S2 and Fig.~S5). In practice, to determine the direction we want to go, we measure the position of the ball $(x,y)$ at a given time (Fig.~S4) and compare it with the position of the next checkpoint on the trajectory, which we try to reach up to a certain threshold distance before moving on to the next checkpoint.

\textbf{Data availability}
The data underlying the figures in this study will be made available in an online repository before the official publication date of the paper.

\textbf{Code availability}
The codes that support the findings of this study are available from the authors upon reasonable request.

\textbf{Supplementary information}
Supplementary information, available at \textcolor{red}{XX}, contains additional information on the scattering matrices extraction (Section 1.1), signal processing and wave state excitation (Section 1.2), regularization method for the measured scattering matrices (Section 1.3), the detection and localization of the target (Section 1.4), the construction of Wigner-Smith operator and the $\boldsymbol{S}$ gradient approximation (Section 1.5), the angular dependence of the gradient approximation accuracy (Section 1.6), and the target manipulation algorithm and the control design (Section 1.7).


\textbf{Acknowledgements}
This work was supported by the Swiss National Science Foundation under the grant SPARK CRSK-2\_190728, by internal EPFL funding, as well as by Nazarbayev University under the Faculty-development competitive research grants program for 2022–2024 Grants No 11022021FD2901 B.O.

\textbf{Contributions} 
N.B., S.R, and R.F. proposed the idea, raised the funding, and supervised the project. B.O. built the experimental set-up, devised the methods, and wrote the codes. B.O. performed the wave-momentum transfer experiments with M.M.  B.O. and M.M. wrote the methods and the supplementary information. B.O., M.M. and R.F. wrote the initial draft of the manuscript. All authors contributed to the writing of the final draft and critically discussed the results.

\textbf{Competing interests}
The authors declare no competing interests.


\bibliography{mainbib}

\end{document}